\documentclass[journal=jacsat,manuscript=article]{achemso}
\setkeys{acs}{articletitle=true}
\usepackage[utf8]{inputenc}
\usepackage{geometry}

\usepackage{amsmath}
\usepackage{amsfonts}
\usepackage{amssymb}
\usepackage{amsthm}
\usepackage{mathtools} 
\usepackage{bm}        
\usepackage{upgreek}   

\usepackage{graphicx}
\usepackage[center]{subfigure} 
\usepackage{subcaption}
\usepackage{adjustbox}
\usepackage{booktabs}
\usepackage{multirow}
\usepackage{xcolor}

\usepackage{achemso}   
\usepackage{comment}   
\usepackage{xparse}    

\usepackage{xr-hyper}
\usepackage[hidelinks]{hyperref}

\makeatletter

\newlength\tindent
\newcommand*{\diff}{\mathop{}\!\mathrm{d}}

\newcommand{\tr}[1]{\text{Tr }{#1}}

\newcommand\epsilonxc{\varepsilon\ped{xc}}
\newcommand\epsilonx{\varepsilon\ped{x}}
\newcommand\epsilonc{\varepsilon\ped{c}}

\NewDocumentCommand{\bh}{G{}G{}}{\tr{\mathbf{h}^{#1}\mathbf{D}^{#2}}}
\NewDocumentCommand{\bJ}{G{}G{}}{\tr{\mathbf{D}^{#1}\mathbf{J}(\mathbf{D}^{#2})}}
\NewDocumentCommand{\bK}{G{}G{}}{\tr{\mathbf{D}^{#1}\mathbf{K}(\mathbf{D}^{#2})}}
\NewDocumentCommand{\bex}{G{}G{}}{\int \rho^{#1}(\mathbf{r})\epsilonx(\rho^{#2}(\mathbf{r}))\diff \mathbf{r}}
\NewDocumentCommand{\bec}{G{}G{}}{\int \rho^{#1}(\mathbf{r})\epsilonc(\rho^{#2}(\mathbf{r}))\diff \mathbf{r}}
\NewDocumentCommand{\bexc}{G{}G{}}{\int \rho^{#1}(\mathbf{r})\epsilonxc(\rho^{#2}(\mathbf{r}))\diff \mathbf{r}}
\NewDocumentCommand{\bvxc}{G{}G{}}{\int v\ped{xc}(\rho^{#1}(\mathbf{r}))\chi_\mu(\mathbf{r})\chi_\nu(\mathbf{r})\diff \mathbf{r}}
\NewDocumentCommand{\bvx}{G{}G{}}{\int v\ped{x}(\rho^{#1}(\mathbf{r}))\chi_\mu(\mathbf{r})\chi_\nu(\mathbf{r})\diff \mathbf{r}}
\NewDocumentCommand{\bvc}{G{}G{}}{\int v\ped{c}(\rho^{#1}(\mathbf{r}))\chi_\mu(\mathbf{r})\chi_\nu(\mathbf{r})\diff \mathbf{r}}

\newcommand\ped[1]{\ensuremath{_{\text{#1}}}}

\graphicspath{{images/}} 

\author{Sara Gómez}
\affiliation{Universidad Nacional de Colombia, Departamento de Química, Av. Cra 30 45-03, 111321, Bogotá, Colombia}
\email{sagomezam@unal.edu.co}
\author{Usman Ali}
\affiliation{Department of Physics, University of Rome Tor Vergata, Via della Ricerca Scientifica 1, 00133, Rome, Italy}
\author{Alessia Muroni}
\affiliation{Department of Physics, University of Rome Tor Vergata, Via della Ricerca Scientifica 1, 00133, Rome, Italy}
\author{Andrea Mele}
\affiliation{Department of Chemistry, Materials and Chemical Engineering ``Giulio Natta'', Politecnico di Milano, Milano, 20133, Italy}
\author{Maria Enrica Di Pietro}
\affiliation{Department of Chemistry, Materials and Chemical Engineering ``Giulio Natta'', Politecnico di Milano, Milano, 20133, Italy}
\author{Tommaso Giovannini}
\affiliation{Department of Physics, University of Rome Tor Vergata, Via della Ricerca Scientifica 1, 00133, Rome, Italy}
\email{tommaso.giovannini@uniroma2.it}

\title[]{Unveiling the Molecular Driving Forces of Pollutant Extraction by Hydrophobic Eutectic Solvents}

\begin{document}

\begin{abstract}
    Hydrophobic eutectic solvents (HES) are emerging as sustainable alternatives to conventional organic solvents for the extraction of molecular pollutants from water. Yet, their selectivity remains poorly understood, hindering the predictive design of eutectic solvents beyond empirical success. Here, we present a multiscale strategy to rationalize and predict solute partitioning in HES. Focusing on bisphenol A (BPA) in trioctylphosphine oxide (TOPO):menthol as a prototypical system, we combine monophasic and biphasic molecular dynamics with quantum energy decomposition of dominant solvation motifs. Our methodology captures the experimentally measured BPA spontaneous migration and thermodynamic stabilization in the HES phase but also identifies the microscopic origin of selectivity: cooperative hydrogen bonding couples to strong dispersion and polarization in the hydrophobic eutectic microenvironment. The robustness of our workflow paves the way for the predictive in-silico screening and design of HES formulations for green and sustainable applications.
\end{abstract}

\section{Introduction}

The replacement of conventional organic solvents with more sustainable alternatives represents a central challenge in modern green chemistry, particularly in the context of separation processes and water remediation.\cite{clarke2018green,welton2015solvents,schuur2019green} Among emerging contaminants, bisphenol A (BPA) stands out as a paradigmatic endocrine-disrupting compound, whose widespread use and persistence in water environments call for efficient, selective, and environmentally benign removal strategies.\cite{mishra2023bisphenol,arnold2013relevance,belfroid2002occurrence,godiya2022removal,tarafdar2022hazardous,ighalo2024bisphenol} In this framework, hydrophobic eutectic solvents (HES) have recently emerged as a highly promising class of extraction media,\cite{van2015hydrophobic,van2020curious} combining low volatility, tunable composition, and favorable sustainability metrics with excellent extraction efficiencies.\cite{cao2021hydrophobic,florindo2020hydrophobic,rodriguez2021extractive,carotti2025hydrophobic} Recent experimental studies have demonstrated that properly designed HES can outperform traditional solvents in the liquid–liquid extraction of BPA from water, establishing these systems as viable candidates for green separation technologies.\cite{schincaglia2025eutectic,ma2024deep}

From a broader perspective, however, the rational development and optimization of sustainable solvents remains a major open challenge. The compositional flexibility of HES gives access to an enormous chemical space,\cite{abbott2003novel,abbott2004deep} which cannot be explored efficiently through purely experimental trial-and-error approaches. As a result, most current HES formulations are discovered through empirical screening, with limited transferability across different solutes and separation targets. In this context, computational screening emerges as a crucial tool to guide and accelerate experimental investigations, reduce material consumption, and facilitate a more rational design of green solvents.\cite{odegova2024designsolvents,escobedo2025rational,sheldon2018metrics,bragagnolo2025overview,kim2025computational,sepali2024deciphering} Crucially, such screening strategies must be able not only to rank solvents based on macroscopic observables, but also to uncover the molecular origins of solute selectivity. The preferential solvation of a molecule in a specific solvent is a collective phenomenon that arises from a delicate interplay between enthalpic and entropic contributions,\cite{ben2013solvation} from the dynamic organization of the solvent environment around the solute,\cite{marcus2013preferential,lum1999hydrophobicity,levy1998computer} and from the strengths of the solute-solvent specific interactions.\cite{sapt1,sapt2,giovannini2020molecular,giovannini2025energy,giovannini2025modeling} Capturing these effects requires approaches that go beyond static structural models and explicitly account for the dynamical nature of solvation in complex condensed-phase systems. 

In this work, we focus on the trioctylphosphine oxide (TOPO):menthol HES, which has been experimentally identified as a particularly efficient solvent for BPA extraction from water\cite{carotti2025hydrophobic}, and therefore represents an ideal model system to elucidate the molecular origins of solvent selectivity. The BPA–HES interaction is characterized by complex and heterogeneous networks of non-covalent interactions, involving multiple and competing hydrogen-bonding motifs as well as significant dispersion contributions arising from the hydrophobic components of the eutectic mixture. Such interaction patterns are intrinsically dynamic and cannot be rationalized on the basis of simple structural arguments alone. To address this challenge, we adopt an integrated theoretical–computational strategy that combines advanced molecular dynamics simulations with high-level quantum mechanical energy decomposition analyses. Molecular dynamics enables us to capture the full dynamical landscape of BPA solvation and to identify how preferential solvation emerges from proper sampling of the phase space, while quantum-level analyses of representative configurations provide a quantitative dissection of the underlying interaction mechanisms. Indeed, molecular dynamics has emerged as a powerful tool to characterize the stability mechanism of HES\cite{pour2022structure, paul2023stability}, the interfacial structure, hydrogen bonding, and phase behavior of HES and water mixtures \cite{paul2021phase,salehi2021interfacial}, highlighting key phenomena of solvation and segregation.\cite{paul2020molecular,fan2021preparation,paul2022study,paul2023decontamination,singh2024investigate} 
By coupling it with quantum-based energy decomposition, we unravel the balance between electrostatics, polarization, exchange–repulsion, and dispersion. Our hybrid dynamic-static approach thus reveals the molecular driving forces responsible for the enhanced stabilization of BPA in the HES relative to water. Beyond rationalizing existing experimental observations, our results establish a robust and transferable framework for the theory-guided screening of sustainable solvents.

\section{Results and Discussion}

In this section, we present and discuss results from two complementary, multiscale perspectives to rationalize the experimental behavior of BPA. First, a molecular dynamics (MD) framework is used to characterize BPA solvation and transfer, combining monophasic simulations in water and in HES with explicit biphasic simulations that directly capture phase separation and pollutant migration (see also Methods). Structural descriptors and solvation free energies from monophasic systems are complemented by potential of mean force calculations that quantify the free-energy cost of transferring BPA across the water–HES interface. Second, a quantum mechanical (QM) approach is employed to resolve the local interaction physics driving these trends, by applying energy decomposition analysis to statistically representative solute–solvent motifs extracted from the MD trajectories. Together, these approaches provide a coherent molecular-level explanation for the preferential stabilization of BPA in HES and the thermodynamic driving force underlying its efficient extraction.

\subsection{A Molecular Dynamics approach to the problem}

\subsubsection{Monophasic solvation of BPA in water vs HES}

\begin{figure}
    \centering
    \includegraphics[width=0.85\linewidth]{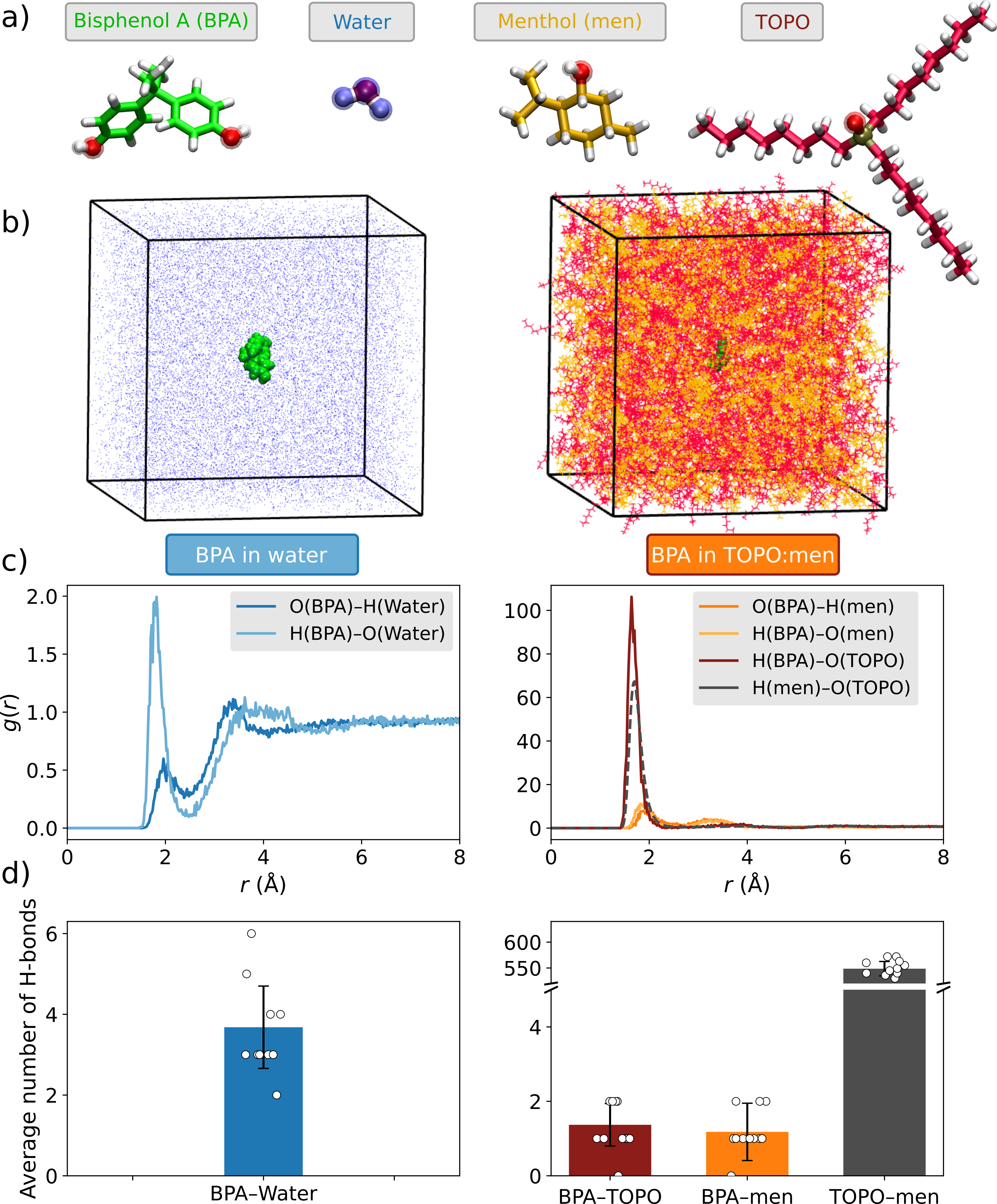}
    \caption{(a) Molecular structures of BPA, water, menthol (men), and TOPO with hydrogen-bond donor/acceptor sites highlighted. (b) Simulation boxes for monophasic BPA in water (left) and in HES (TOPO:men = 1:2, right). (c) Radial distribution functions $g(r)$ for BPA–solvent interactions in water and HES (solid lines: BPA–TOPO/BPA–menthol; dashed: TOPO–menthol). (d) Average number of hydrogen bonds and standard deviations for BPA–water and BPA–HES interactions, with subsampled trajectory values (every 300 frames) shown as points.}
    \label{fig:monophasic}
\end{figure}

Before addressing phase separation and extraction-related phenomena, we first establish the intrinsic solvation preferences of BPA in homogeneous environments. 30 ns monophasic MD simulations were therefore employed to represent the individual BPA-solvent interactions in the condensed phase in TOPO:men and in water separately, to capture cooperative H-bonding from the sampling (see also Methods). Fig. \ref{fig:monophasic}a-b graphically reports the molecular structures of the monophasic components and provides a representation of the simulation boxes exploited in the MD runs (left, water; right HES). The TOPO:men MD was set to match the molar ratio used in the experiments (1:2).\cite{carotti2025hydrophobic} As an initial validation, the computed density of the TOPO:men phase was 986.4 kg m$^{-3}$, differing by 124.2 kg m$^{-3}$ from liquid water and in close agreement with the reported experimental value of 915.7 $\pm$ 8.9 kg/m$^{-3}$.\cite{carotti2025hydrophobic}  Time evolution of the densities is reported in Figure S1 of the Supporting Information (SI). 
The calculated self-diffusion coefficients in the monophasic HES system follow the expected mobility hierarchy ($D_{\rm men} > D_{\rm TOPO}  > D_{\rm BPA}$), with values of 2.55 $\times 10^{-11}$, 1.12 $\times 10^{-11}$, and 1.05 $\times 10^{-11}$ m$^2$s$^{-2}$, respectively. A rough viscosity estimate derived from the Stokes–Einstein relation using the diffusion coefficients of menthol and TOPO (assuming reasonable hydrodynamic radii of 0.35–0.40 nm and 0.6–0.8 nm, respectively) yields $\eta \approx$ 20–60 mPa$\cdot$s and 24–32 mPa$\cdot$s, both consistent with the experimental value of 41.47 ± 0.94 mPa$\cdot$s. Although the Stokes–Einstein approximation neglects molecular anisotropy and specific interactions, the agreement supports the realistic dynamic behavior of the simulated HES phase. 

The local solvation structure of BPA was first examined through radial distribution functions (RDFs), which also served to define the geometric criteria used in the hydrogen-bond analysis. Figure \ref{fig:monophasic}c shows RDFs corresponding to hydrogen-bond–relevant atom pairs between BPA and its surrounding environment. In water, BPA–water RDFs exhibit common first-shell maxima at $\sim$1.9~\AA, dominated by interactions between BPA oxygen atoms and water hydrogens. These features are consistent with transient, rapidly exchanging hydrogen bonds typical of aqueous solvation.\cite{zhang2016molecular,wernet2004structure,liu2018hydrogen}

In contrast, BPA–HES RDFs display sharper and much more intense first peaks in the 1.8–2.2~\AA\ range, indicative of stronger and more directional interactions. In particular, the P=O $\cdots$ H–O(BPA) correlation associated with TOPO exhibits a pronounced and narrow first-shell peak, revealing a well-defined local solvation structure in which TOPO molecules preferentially coordinate BPA. Interactions involving menthol are also present but exhibit broader and less intense peaks, reflecting weaker and more flexible hydrogen bonding. This distinction can be rationalized by the strongly polarized and sterically exposed P=O moiety of TOPO, which acts as a highly effective hydrogen-bond acceptor, whereas the menthol hydroxyl group is less accessible due to its bulky hydrophobic framework. Notably, the extent of the TOPO–menthol RDFs highlights the cooperative organization of the HES microstructure, which provides a structured environment that neither component could achieve independently.

While RDFs provide insight into local structure, hydrogen-bond statistics offer a quantitative comparison between solvents. Figure \ref{fig:monophasic}d reports the average number of hydrogen bonds formed by BPA in water and in HES. In monophasic simulations, BPA forms on average 3.7 hydrogen bonds with water, compared to 1.4 and 1.2 hydrogen bonds with TOPO and menthol, respectively. Individual data points, subsampled from the trajectories, reveal substantial temporal fluctuations, underscoring the dynamic nature of hydrogen bonding in both environments. The reduced number of BPA–HES hydrogen bonds relative to water does not imply weaker solvation; rather, it reflects fewer but more persistent and directional interactions. The internal hydrogen-bond network of the HES is dominated by TOPO–menthol interactions, with an average of approximately 585 hydrogen bonds. This extensive network remarkably surpasses solute–solvent interactions and establishes TOPO–menthol as the structural backbone of the eutectic phase. In contrast, TOPO–TOPO hydrogen bonds are rare, and menthol–menthol interactions are primarily dispersive with occasional OH–OH contacts. The hydrogen-bond structure of bulk water is well reproduced, providing a reliable reference for BPA solvation.

To probe interaction dynamics, hydrogen-bond lifetimes ($\tau$) were evaluated by using both continuous and intermittent definitions.\cite{luzar1996hydrogen,luzar2000resolving} BPA–water hydrogen bonds exhibit short lifetimes, on the order of a few picoseconds, consistent with rapid exchange and diffusive solvation. BPA–menthol interactions display intermediate residence times ($\sim$60~ps), and fully decorrelate within the simulation time considered in this work (30 ns). Notably, as a projection, BPA–TOPO hydrogen bonds persist for hundreds of ns or longer, effectively spanning the entire trajectory and excluding interfacial adsorption or kinetic trapping. The outlined hierarchy of lifetimes, $\tau_{\rm BPA-TOPO} \gg \tau_{\rm BPA-menthol} \gg \tau_{\rm BPA-water}$, reveals a cooperative mechanistic picture: TOPO acts as the primary hydrogen bonding site for BPA, while menthol provides a dynamic yet stabilizing local environment.

Consistent with this observation, minimum-distance analyses show that menthol remains within the first BPA solvation shell ($<0.27$~nm). This indicates that menthol is indeed present around the solute along the whole trajectory, modulating and determining the strength of BPA–TOPO interactions. Together, these results demonstrate that BPA is inherently more stably solvated in HES than in water. Remarkably, this is not the result of a higher number of contacts, but of fewer, but stronger, and longer (in time) interactions. This is one of the main findings of our work: through monophasic simulations, the thermodynamic and kinetic foundations for BPA extraction can be established, independently of phase-separation effects.

The preferential solvation of BPA in HES rather than water, suggested by the structural descriptors discussed above, can also be quantified thermodynamically by comparing the free energy cost of separating BPA from its solvent environment in water and in HES. In particular, we characterized BPA solvation using potentials of mean force (PMFs), which directly report the reversible work required to separate BPA from its surrounding solvent along a physically meaningful coordinate. 

The resulting monophasic PMFs in water and in HES, together with the corresponding transfer free energy profile ($\Delta G$), are shown in Figure S2 of the SI. In water, the PMF starts close to 0 kcal/mol at short separation (0.20 nm), decreases almost monotonically, and reaches approximately $\Delta G =$ -7.0 $\pm$ 0.5 kcal/mol at about 0.9 nm. The free-energy minimum is located at 0.45 nm, consistent with a water molecule occupying the first hydration shell of BPA. This is well in agreement with the sharp first-shell features observed in the BPA–water RDFs (Figure \ref{fig:monophasic}c). For the BPA-HES system, the PMF minimum is deeper, reaching -11.4 kcal/mol at a separation of $\sim$ 1.10 nm. The shift of the minimum to larger distances reflects the larger molecular size and extended contact surface of the TOPO and menthol components, rather than a fundamentally different interaction mechanism. In both solvents, the negative PMF values indicate thermodynamic stabilization of BPA by solvation, with magnitudes characteristic of moderate solvation free energies for solutes containing hydrophobic parts as BPA, rather than strong, site-specific binding.

A direct comparison of the two PMFs through the distance-dependent transfer free energy, $\Delta G_{transfer} = \Delta G^{HES}_{binding} - \Delta G^{water}_{binding}$ (see Fig. S2 in the SI), reveals that BPA experiences comparable stabilization in water and HES at short and intermediate separations, but exhibits a substantial thermodynamic preference for the HES at larger separations ($\Delta G =$ -11.4 kcal/mol when distance $>$ 0.9 nm). This behavior is fully consistent with the previous analysis, demonstrating that preferential solvation in HES does not arise from solute-solvent dominant interactions, but from the cumulative stabilization provided by a heterogeneous hydrophobic environment\cite{gomez2021molecular} that can maintain favorable BPA–solvent contacts over time.

\subsubsection{Biphasic simulations: spontaneous partitioning and phase separation}

\begin{figure}[!htbp]
    \centering
    \includegraphics[width=0.5\linewidth]{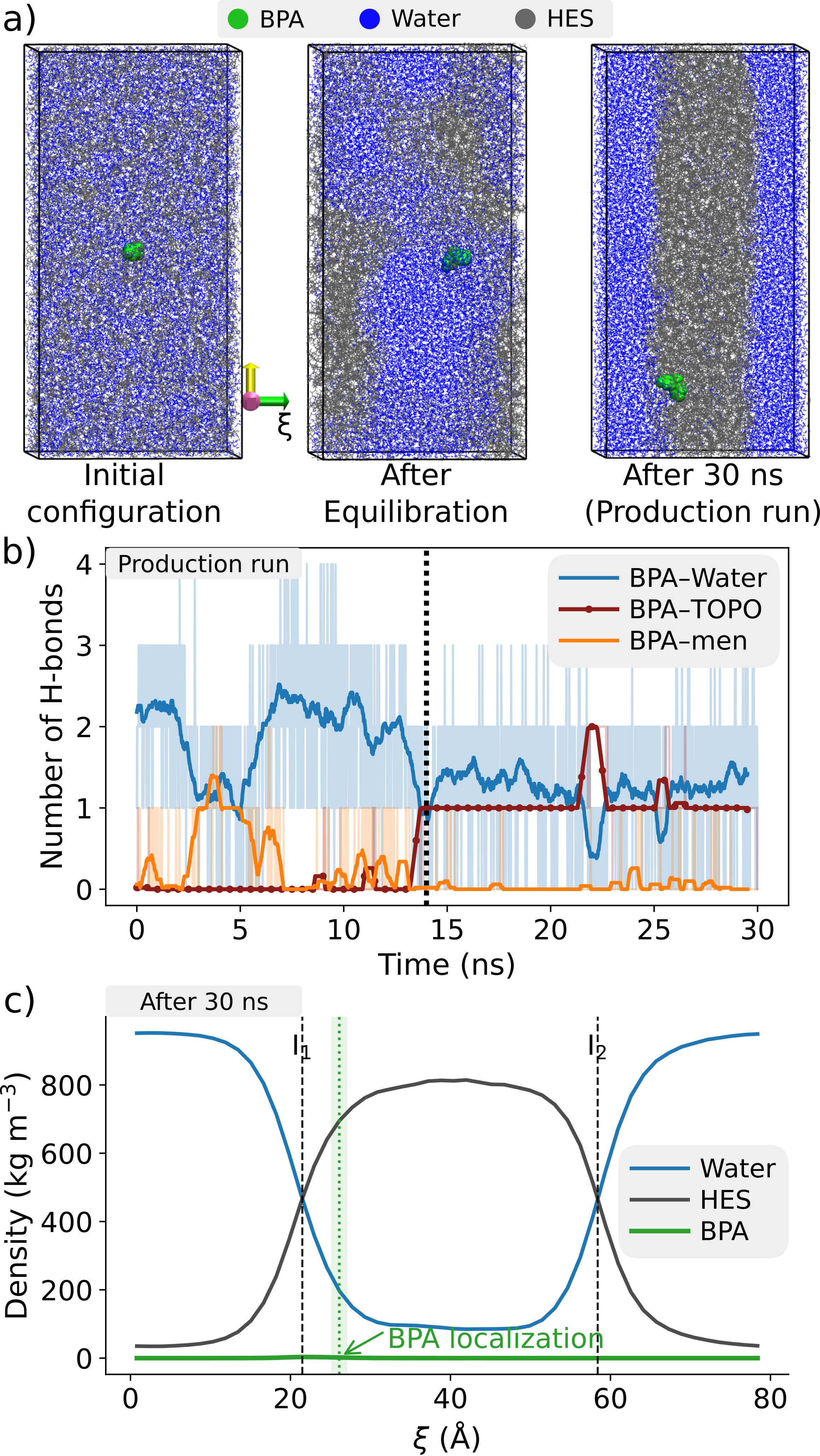}
    \caption{(a) Time evolution of the biphasic water–HES system containing BPA from the initial fully mixed configuration (left) to the phase-separated state at $t=30$ ns (right). (b) Time evolution of BPA hydrogen bonds with water, TOPO, and menthol during the production run (30 ns); the vertical dashed line at $t=14$ ns marks BPA transfer from water to HES. (c) Density profiles along $\xi$ coordinate at the end of the production run (panel a), showing the preferential accumulation of BPA within the HES-rich region at $\xi \approx 26.1$~\AA\ (green arrow); dashed lines indicate the water–HES interfaces (I$_1$, I$_2$) arising from periodic boundary conditions.}
    \label{f:biphasic}
\end{figure}

Although monophasic simulations can establish the intrinsic solvation preferences of BPA in water and in HES, extraction is inherently a biphasic process that involves interfacial reorganization and solvent competition. To directly capture this mechanism at the molecular level, we performed atomistic biphasic MD simulations in which water and HES phases coexist, so that BPA can redistribute freely between them. In this way, we provide a simultaneous characterization of phase separation, interfacial structure, hydrogen-bonding dynamics, density organization, and BPA migration, thereby achieving a direct link between solvation thermodynamics and the experimentally observed extraction behavior reported in Ref. \citenum{carotti2025hydrophobic}.

Figure \ref{f:biphasic} summarizes the structural evolution, hydrogen-bond dynamics, and spatial organization of a biphasic HES–water system containing a single BPA molecule. The sequence of snapshots shown in Figure \ref{f:biphasic}a illustrates the spontaneous evolution from an initially mixed configuration toward a fully phase-separated system. During equilibration, water and HES rapidly segregate into distinct domains, while BPA migrates from the mixed region into the HES-dominated phase. Importantly, this redistribution occurs without external bias and emerges naturally from the interatomic interactions described by the classical force fields (TIP3P\cite{mark2001structure_tip3p}, GAFF2\cite{he2020fast_gaff2}). Indeed, the final snapshot used to initiate the production stage reveals that BPA is fully embedded within the bulk HES phase and it is not localized at the interface. This indicates a genuine phase partitioning rather than transient interfacial adsorption.

Molecular-level insight into the origin of BPA stabilization in the biphasic environment is provided by the RDFs reported in Figure S4 in the SI. Even in the presence of water, solute–solvent RDFs confirm the persistence of strong, short-range BPA–HES interactions, which mainly involve the TOPO P=O group. At the same time, solvent–solvent RDFs demonstrate that the intrinsic hydrogen-bond networks of bulk water and bulk HES remain largely unperturbed. These results indicate that BPA insertion into the HES phase does not alter the solvent structure. Instead, BPA is stabilized by the creation of additional BPA-HES hydrogen-bond frameworks, consistent with the monophasic solvation analysis. The dynamical evolution of hydrogen bonding further clarifies the mechanism of selective partitioning. Figure \ref{f:biphasic}b shows the time-dependent number of HBs formed between BPA and each solvent component (water, TOPO, and menthol) throughout the trajectory. In the early stages, when the system is still partially mixed (see Fig. \ref{f:biphasic}a middle panel), BPA interacts predominantly with water, intermittently exchanging hydrogen bonds with menthol to maintain a fluctuating network of approximately two hydrogen bonds. As phase separation progresses, BPA–water hydrogen bonds decrease sharply, while a pronounced increase in BPA–TOPO hydrogen bonding is observed in the late stages of the simulation ($t \gtrsim 14$ ns). This transition coincides temporally with BPA migration into the HES-rich domain and mirrors the hierarchy of hydrogen-bond lifetimes identified in the monophasic simulations, confirming the dominant role of BPA–TOPO interactions also in the biphasic trajectory.

An additional confirmation of phase separation and BPA localization within the HES is obtained from the density profiles along the $\xi$ coordinate (aligned with the simulation box $y$-axis), shown in Figure \ref{f:biphasic}c. The total system density averages to 938.3 kg m$^{-3}$, in good agreement with the experimental value of 915.7 $\pm$ 8.9 kg m$^{-3}$.\cite{carotti2025hydrophobic}. The individual density profiles of water and HES intersect at two well-defined interfaces, $I_1$ and $I_2$, located at 21.7 \AA{} and 59.4 \AA{}, respectively, confirming the formation of stable, macroscopically separated phases. In contrast, BPA exhibits a single, broad density maximum centered at $\xi \approx 26.1$ \AA, well within the HES region, with negligible density in the aqueous phase. Remarkably, no significant BPA density is observed at the interfaces, demonstrating the suggested interfacial trapping as the dominant extraction mechanism.

Our biphasic simulations demonstrate that selective partitioning of BPA into the HES phase arises spontaneously from solvent competition and hydrogen-bond cooperativity, overcoming interfacial adsorption or kinetic trapping. Having established the structural and dynamical origin of BPA migration, we next quantify the thermodynamic driving force for transfer across the water–HES interface.

\begin{figure}
    \centering
    \includegraphics[width=0.7\linewidth]{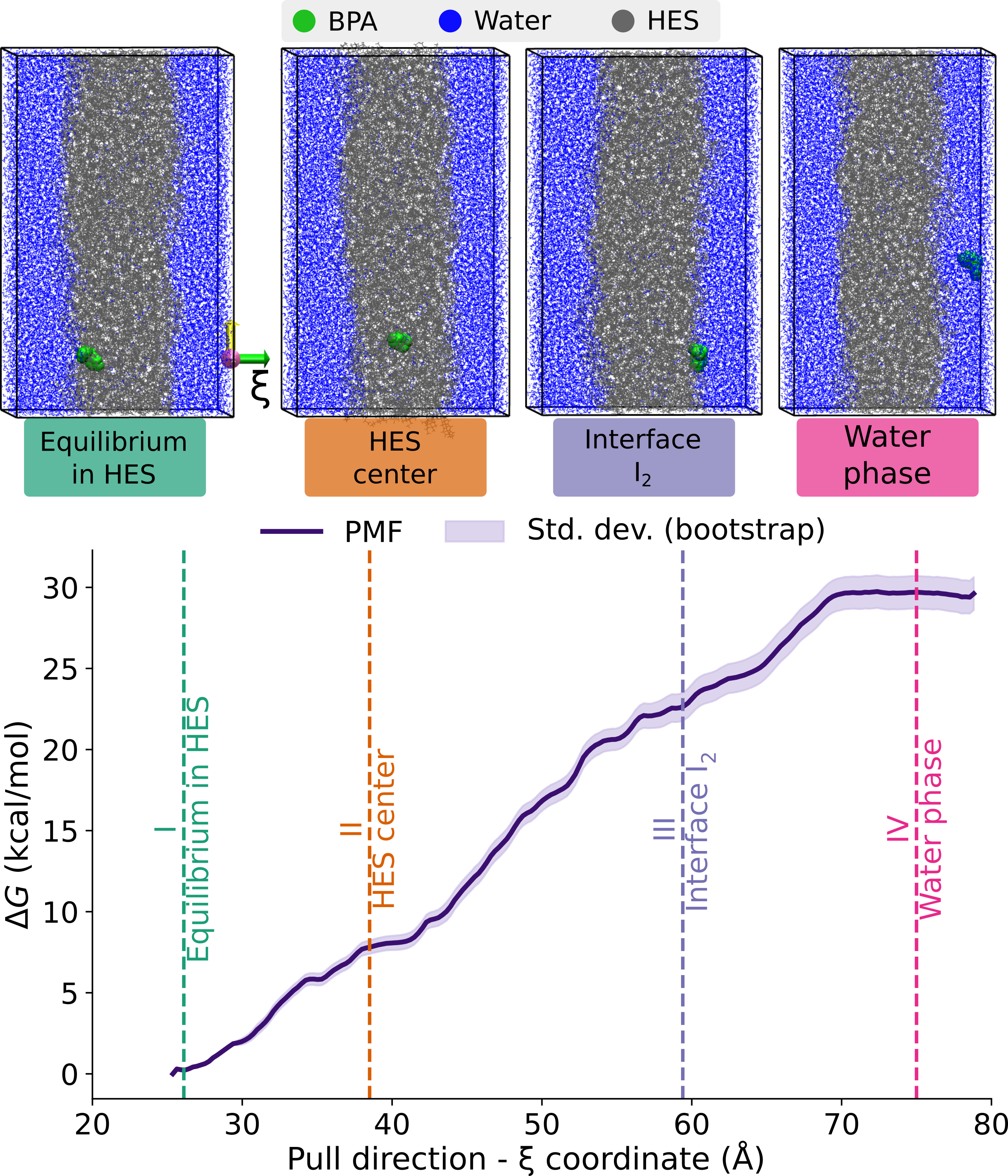}
    \caption{Free-energy profile of BPA transfer from HES to water obtained from umbrella sampling simulations as a function of the pull direction $\xi$, shown as the mean Potential of Mean Force (PMF, solid line) with bootstrap standard deviation (shaded region). Vertical lines highlight representative BPA positions along the pathway (equilibrium in HES, center of HES domain, HES–water interface, and water phase).}
    \label{f:pmf_bpa_transfer}
\end{figure}

To connect the intrinsic solvation thermodynamics identified in the monophasic reference simulations (Figure \ref{fig:monophasic}) with the molecular mechanism of extraction under biphasic conditions, we quantified the free-energy cost associated with transferring BPA from the HES phase into water across the interface. Figure \ref{f:pmf_bpa_transfer} reports the PMF describing this transfer process along the interface normal. Representative configurations sampled along the reaction coordinate, which correspond to BPA equilibrated in HES (I), migration through the interfacial region (II and III), and final solvation in water (IV), are shown in Figure \ref{f:pmf_bpa_transfer} (top panel) to aid interpretation of the free-energy landscape.

The PMF obtained from steered MD, umbrella sampling, and WHAM analysis (See Methods) is shown in Figure \ref{f:pmf_bpa_transfer} (bottom panel), where bootstrap-derived standard deviations are also given. The error percentages remain small throughout the profile, indicating good statistical convergence. As can be appreciated, the free-energy minimum is reported for BPA dissolved in the HES-rich region, i.e., for the equilibrium BPA position identified in the unbiased biphasic MD ($\xi = 26.1$ \AA{}, green line). As BPA is moved toward the aqueous phase, the free energy increases monotonically. In particular, no secondary minima at either interface are found, confirming that the water–HES interface does not provide thermodynamic stabilization. The overall free-energy penalty for transferring BPA from HES into bulk water is 29.5 $\pm$ 1.0 kcal/mol, far exceeding thermal fluctuations at ambient conditions and establishing a strong thermodynamic preference for BPA retention in the eutectic phase.

Together with the structural and dynamical evidence presented above, this free-energy profile demonstrates that selective partitioning of BPA into HES is governed by bulk solvation thermodynamics. Having established the macroscopic driving force for extraction, we next examine the microscopic origin of this stabilization by resolving the specific intermolecular interactions responsible for BPA binding within the HES microenvironment using quantum chemical analysis.

\subsection{A Quantum Mechanical approach to the problem}

To validate the free-energy analysis obtained from classical MD simulations and to provide a robust quantum-mechanical reference for the interaction of BPA with the two solvents, we investigated the nature of solute–solvent interactions using Kohn–Sham Fragment Energy Decomposition Analysis (KS-FEDA).\cite{giovannini2024kohn,giovannini2025energy} This approach relies on the variational minimization of the energies of the individual fragments within the electronic structure of the full adduct.\cite{giovannini2021energy,giovannini2022fragment} As a result, the method yields fragment-localized orbitals optimally adapted to the interacting environment, enabling a physically transparent and quantitatively reliable decomposition of interaction energies. When coupled with the D4 dispersion correction, KS-FEDA can reproduce high-level, golden standard, Symmetry-Adapted Perturbation Theory (SAPT) reference data,\cite{sapt1,sapt2,giovannini2024kohn} with errors well below chemical accuracy ($<$ 1 kcal/mol), while retaining a reduced computational cost. 

\begin{figure}[!htbp]
    \centering
    \includegraphics[width=1\linewidth]{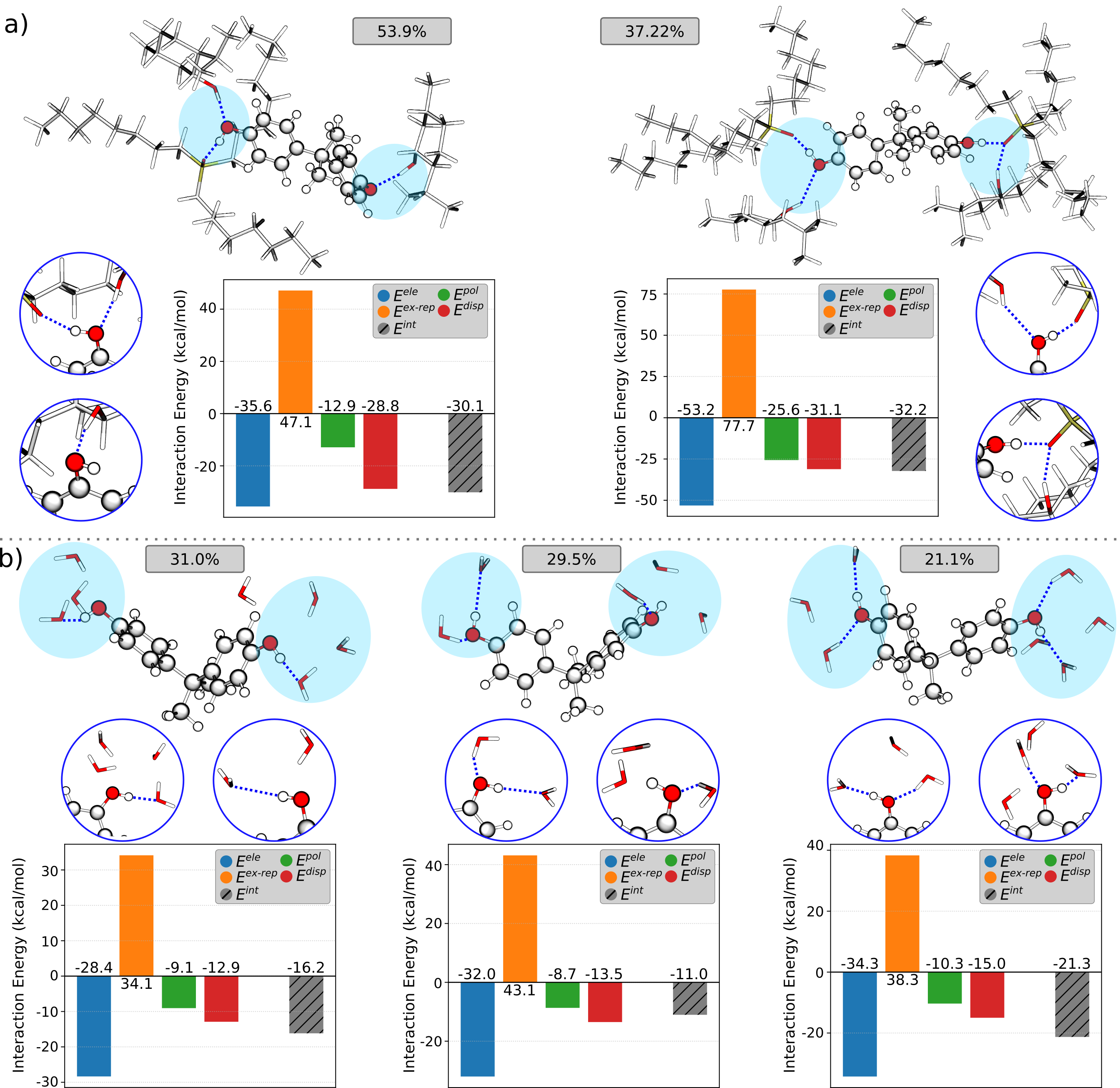}
    \caption{(a-b, top) Graphical depiction of representative structures of BPA dissolved in HES (a) and in water (b). (a-b, bottom) KS-FEDA energetic terms in kcal/mol. The insets highlight the solute-solvent specific interactions (Hydrogen Bonding) for each cluster.}
    \label{fig:ks-feda}
\end{figure}

KS-FEDA is here applied to a set of representative structures describing BPA solvated in HES and in water, which are extracted from the MD trajectories via a clustering procedure (see Methods) and are representative of the dominant solvation motifs. Considering the computational cost associated with a QM description, the interaction analysis is restricted to the first solvation shell as outlined by the RDFs depicted in Fig. \ref{fig:monophasic}c, defined by a cutoff distance of 3.8 \AA~ for the HES and 3.5 \AA{} for water. The representative configurations with populations larger than 20\% are shown in Fig. \ref{fig:ks-feda}a–b (top), where the dominant non-covalent interactions are highlighted. In the case of BPA–HES, the two depicted structures account for approximately 90\% of the total population, and are characterized by BPA acting simultaneously as a hydrogen-bond acceptor toward menthol and as a hydrogen-bond donor toward the P=O group of TOPO. In the second motif (Fig. \ref{fig:ks-feda}a, right), an additional bridging arrangement involving both TOPO and menthol is observed. Similarly, the three representative water configurations (total population $\sim$ 80\%) display qualitatively similar hydrogen-bonding patterns between BPA and surrounding water molecules, highlighting that a purely geometric inspection of the solvation structures does not, by itself, provide a clear qualitative rationale for the stronger affinity of BPA toward the HES compared to water. This analysis is perfectly consistent with the results obtained from monophasic and biphasic MDs discussed above.

KS-FEDA decomposes the total interaction energy, $E^{int}$, into electrostatic ($E^{ele}$), polarization ($E^{pol}$), exchange–repulsion ($E^{ex-rep}$), and dispersion ($E^{disp}$) contributions (see Methods). Polarization and dispersion are attractive terms, whereas the exchange–repulsion contribution is positive overall, despite containing an intrinsically attractive quantum-mechanical exchange component. The electrostatic contribution may be either stabilizing or destabilizing, depending on the relative orientation and complementarity of the static multipoles. The KS-FEDA energy terms for the representative structures are graphically depicted in Fig. \ref{fig:ks-feda}a-b, bottom panels (see Tabs. S3-S4 in the SI for raw data). For the BPA–HES adducts, the interaction energies are markedly stabilizing, with $E^{int}$ values of -30.1 and -32.2 kcal/mol for the two dominant structures. Such an enhanced stabilization arises from a synergistic interplay of electrostatics, polarization, and dispersion. In particular, the electrostatic term is significantly favorable (-35.6 and -53.2 kcal/mol), reflecting the strong directional hydrogen bonds involving both menthol and TOPO. Polarization is also relevant (-12.9 and -25.6 kcal/mol), consistent with the high polarizability of the HES components. Most notably, dispersion provides a large stabilizing contribution (-28.8 and -31.1 kcal/mol), underscoring the importance of non-specific, long-range interactions between BPA and the hydrophobic solvent molecules. Although the exchange–repulsion term is correspondingly large in magnitude, it does not compensate for the strong attractive contributions, resulting in a deeply stabilized adduct.

In contrast, the BPA–water interactions are significantly weaker, with $E^{int}$ values ranging from -11.0 to -21.3 kcal/mol across the three representative structures. While electrostatics remains stabilizing (-28.4 to -34.3 kcal/mol), both polarization (-8.7 to -10.3 kcal/mol) and dispersion (-12.9 to -15.0 kcal/mol) are considerably reduced compared to the HES case. As a consequence, the balance between attractive and repulsive terms leads to a substantially weaker net interaction. These findings are also confirmed by the average values for each interaction energy term obtained by weighting each cluster value for its population (see Tabs. S1-S2 and Fig. S3 in the SI).

Overall, these results clearly identify dispersion, amplified by enhanced polarization and favorable electrostatics, as the primary driving force behind the preferential stabilization of BPA in the hydrophobic eutectic solvent. Indeed, this also supports the longer BPA-HES hydrogen bonding patterns revealed by the MD analysis. Importantly, average interaction energies computed over all representative structures (about -17 and -30  kcal/mol in water and HES, respectively, see Tabs. S3-S4 in the SI) are also in very good agreement with the trends and magnitudes obtained from the classical MD free-energy analysis (-29.5 kcal/mol for the $-\Delta G_{transfer}$ from the PMF on the biphasic system). Such a consistency demonstrates the robustness of the MD protocol employed here and validates its applicability to complex, heterogeneous solvent environments such as hydrophobic eutectic systems.

\section{Conclusions} \label{sec:sum_conc}

In this work, we have provided a molecular-level elucidation of the preferential solvation of bisphenol A in a prototypical hydrophobic eutectic solvent, TOPO:men, relative to water. By integrating statistically converged molecular dynamics free-energy calculations with quantum-mechanical energy decomposition analysis, we show that solvent selectivity is a genuinely collective and dynamical phenomenon, which cannot be inferred from static structural descriptors alone. Across both homogeneous and biphasic simulations, BPA is consistently stabilized in the HES phase, and the corresponding thermodynamic preference is in line with the experimentally observed extraction behavior. At the microscopic level, this preference does not originate from a larger number of hydrogen bonds, but from cooperative and heterogeneous motifs involving both TOPO and menthol that persist over time and position BPA in a highly polarizable, dispersion-rich microenvironment. Consistently, energy decomposition reveals that, while electrostatics is of comparable magnitude in water and HES, the enhanced stabilization in the eutectic phase is driven by significantly stronger dispersion and polarization contributions.

Beyond rationalizing existing experimental observations, these results establish a general and predictive framework for investigating solute–solvent selectivity in complex condensed-phase systems. The combined use of advanced molecular dynamics and quantum-level energy decomposition enables a direct connection between macroscopic thermodynamic observables and their microscopic interaction origins. This approach is inherently transferable and can be systematically applied to screen and design hydrophobic eutectic solvents with tailored selectivity, thereby guiding experimental efforts toward optimal formulations. More broadly, the methodology presented here paves the way for theory-driven solvent discovery in green chemistry, offering a robust route to replace conventional solvents in separation processes, water remediation, and related applications where molecular-level understanding is essential for rational design.

\section{Methods}

\subsection{Molecular Optimization and Force-Field Parametrization}

Isolated bisphenol A (BPA), trioctylphosphine oxide (TOPO), and menthol (men) molecules were optimized at the quantum mechanical (QM) level to obtain reference geometries. Geometry optimizations and harmonic frequency calculations were carried out using density functional theory (DFT) at the B3LYP/cc-pVDZ level as implemented in Gaussian16 \cite{g16}. All optimized structures were confirmed to be true minima by the absence of imaginary frequencies.

To approximate condensed-phase polarization effects relevant for force-field parametrization, implicit solvation was included via the Polarizable Continuum Model (PCM) \cite{mennucci2012polarizable,tomasi2005quantum}. Dielectric constants of $\varepsilon = 79.5$ and $\varepsilon = 3.2$ were employed to represent water and the hydrophobic eutectic solvent (HES), respectively. The latter value was estimated using the Lichtenecker and linear mixing rules \cite{lichtenecker1924elektrische,lichtenecker1926dielektrizitatskonstante,goncharenko2025century}, based on reported dielectric constants of TOPO ($\varepsilon_r \approx 2.5$--2.6) and menthol ($\varepsilon_r \approx 3.9$--4.0) and their volume fractions at 298 K.

The optimized geometries served as input for classical force-field parametrization using ACPYPE \cite{sousa2012acpype,kagami2023acpype}. Bonded and nonbonded parameters were assigned according to the GAFF2 force field \cite{he2020fast_gaff2}. The resulting parameters were used consistently in all monophasic and biphasic molecular dynamics simulations.

\subsection{Construction of Monophasic and Biphasic Simulation Systems}

Monophasic simulation boxes were constructed for a single BPA molecule solvated in pure water (7348 TIP3P molecules \cite{jorgensen1983comparison,mark2001structure_tip3p}) and for BPA solvated in HES composed of 465 TOPO molecules and 930 menthol (men) molecules, preserving the 1:2 molar ratio (Figure \ref{fig:monophasic}b).

Biphasic systems were generated by initially mixing water (17642 molecules), TOPO (350 molecules), and menthol (700 molecules) in a single simulation box, maintaining the same HES composition used in monophasic simulations. One BPA molecule was placed in the mixed region without imposing any preferential initial location.

\subsection{Molecular Dynamics Simulations}

All molecular dynamics simulations were performed using GROMACS 2020.4 \cite{abraham2015gromacs}. Energy minimization was carried out using the steepest descent algorithm until the maximum force was below 500 kJ mol$^{-1}$nm$^{-1}$. Systems were then equilibrated for 1 ns in the NVT ensemble using the velocity-rescaling thermostat \cite{bussi2007canonical}, followed by 2 ns of NPT equilibration using the Berendsen barostat \cite{berendsen1984molecular} to relax the density. Production simulations of 30 ns were subsequently performed in the NPT ensemble using the Nosé--Hoover thermostat and the Parrinello--Rahman barostat \cite{parrinello1981polymorphic} at 298 K and 1 bar. During production, no restraints were applied, allowing spontaneous phase separation and BPA migration in biphasic systems.

Periodic boundary conditions were applied in all directions. Electrostatic interactions were treated using the Particle Mesh Ewald method \cite{darden1993particle} with a real-space cutoff of 1.0 nm. Van der Waals interactions were truncated at 1.2 nm with long-range corrections. All bonds involving hydrogen atoms were constrained using LINCS \cite{hess1997lincs}, permitting a 2 fs integration time step. Trajectories were saved every 10 ps.

\subsection{Structural and Dynamical Analyses}

Structural and dynamical properties were analyzed using standard GROMACS tools. Radial distribution functions (RDFs) between BPA and solvent species were computed to characterize local solvation environments. Hydrogen bonds were identified using a donor--acceptor distance cutoff of 3.5 \AA{} (or 3.8 \AA{}) and an angular cutoff of 30$^\circ$. Hydrogen-bond lifetimes were computed using both continuous and intermittent correlation functions following the Luzar--Chandler formalism \cite{luzar1996hydrogen,luzar2000resolving}.

Minimum-distance distributions and density profiles were computed to complement the RDF analysis. In biphasic systems, density profiles along the interface normal ($\xi$, aligned with the $y$-axis) were obtained by binning atomic positions and averaging over production trajectories. All reported averages and uncertainties were obtained using block averaging. Representative snapshots were visualized using VMD 1.9.3 \cite{humphrey1996vmd}.

\subsection{Clustering and Extraction of Representative Configurations}

Representative solute--solvent motifs were extracted from monophasic trajectories via RMSD-based clustering using the GROMOS algorithm \cite{daura1999peptide}. Clustering included BPA heavy atoms and first-shell solvent molecules. Five clusters were identified for BPA in water and seven for BPA in HES. Representative structures from the most populated clusters were selected for subsequent quantum mechanical analysis. This clustering-based strategy has been successfully applied to characterize statistically relevant solvation environments in previous studies \cite{gomez2022ring,gomez2023uv,gomez2024modeling,giovannini2025modeling}.

\subsection{Free-Energy Calculations in Monophasic and Biphasic Systems}

Solvation thermodynamics were quantified via potential of mean force (PMF) calculations describing the separation of BPA from its solvent environment. In this way, we avoid the issues related to the calculation of the binding free energies from Molecular Mechanics Generalized Born (Poisson Boltzmann) Surface Area (MM/GB(PB)SA)-type approaches\cite{massova2000combined,kollman2000calculating}, which are commonly used for well-defined host–guest complexes, but are ill-suited for extended solvent environments, where stabilization arises from collective solvation. Additionally, the application of MM/GB(PB)SA would not be suitable for a solute-solvent(s) system, since in those methods, the entropic contributions are poorly defined, and the reference state becomes ambiguous.

Initial configurations were generated using steered molecular dynamics with a pulling velocity of 0.0001 nm ps$^{-1}$ and a force constant of 1000 kJ mol$^{-1}$ nm$^{-2}$. Umbrella sampling simulations were performed using 30--50 windows spaced by 0.02 nm (monophasic) and 0.01 nm (biphasic). Each window was equilibrated for 1 ns, followed by 3 ns of sampling in the NPT ensemble. PMFs were reconstructed using WHAM \cite{hub2010g_wham}, and uncertainties were estimated via bootstrap analysis (100 resamplings).

For biphasic systems, the reaction coordinate was defined as the projection along the $y$ direction of the vector connecting the center of mass of BPA and that of the HES slab. To prevent translational drift of the slab during biased simulations, a weak harmonic restraint was applied to the HES center of mass along the $y$ axis without affecting its internal structure.

An important note regarding the slab stabilization and umbrella sampling is that to prevent translational drift of the HES slab during biased simulations, a weak harmonic restraint was applied along the $y$ direction between the HES center of mass and an internal reference atom located near the slab center. This restraint had zero pulling rate and served exclusively to stabilize the slab position without altering its internal structure.

\subsection{Quantum Mechanical Energy Decomposition Analysis}

Quantum mechanical energy decomposition analyses were performed on representative clusters extracted from monophasic simulations. HES clusters consisted of BPA interacting with 1--2 TOPO and 2 men molecules, while water clusters contained 5--10 water molecules.

Energy Decomposition Analysis (EDA) was performed using the recently developed Kohn–Sham Fragment Energy Decomposition Analysis (KS-FEDA) method,\cite{giovannini2024kohn,giovannini2025energy}, which is rooted in density functional theory (DFT). In KS-FEDA, the interaction energy is obtained through the variational minimization of the electronic energies of the individual fragments within the electronic structure of the full adduct, expressed in terms of Kohn–Sham Fragment Localized Molecular Orbitals (KS-FLMOs).\cite{giovannini2021energy,giovannini2022fragment} This formulation yields fragment orbitals optimally adapted to the interacting environment and enables a physically transparent and quantitatively robust decomposition of intermolecular interactions.\cite{giovannini2024kohn,giovannini2025energy} Among the different decomposition schemes available within KS-FEDA,\cite{giovannini2024kohn} the total interaction energy is here expressed as:

\begin{equation}
E^{int} = E^{ele} + E^{pol} + E^{ex-rep} + E^{disp},
\end{equation}

where $E^{ele}$, $E^{pol}$, $E^{ex-rep}$, and $E^{disp}$ denote the electrostatic, polarization, exchange–repulsion, and dispersion contributions, respectively. 

In this work, KS-FEDA was applied to 5 and 7 representative BPA clusters solvated in HES and in water, respectively, extracted from monophasic MD trajectories using the GROMOS clustering protocol\cite{daura1999peptide}. For each configuration, all solvent molecules within 3.8~\AA\ (HES) and 3.5~\AA\ (water) from any BPA atom were retained in the quantum cluster, thereby explicitly describing the first solvation shell relevant for specific solute–solvent interactions (see also Fig. \ref{fig:monophasic}c). All KS-FEDA calculations were performed at the B3LYP level of theory,\cite{becke1988density,stephens1994ab} and dispersion interactions were treated using the D4 correction.\cite{caldeweyher2017extension,caldeweyher2019generally,caldeweyher2020extension} For BPA-water clusters, all atoms were described using the 6-31+G*, while for BPA–HES clusters, we exploited a mixed basis-set strategy to balance accuracy and computational cost: the atoms involved in potential hydrogen-bonding interactions (i.e. the polar P=O and O–H sites of TOPO and menthol) were described using the 6-31+G* basis set, while the hydrophobic alkyl chains of the eutectic components were treated with the 6-31G basis set. We remark that the chosen level of theory, B3LYP-D4, has been shown to reproduce high-level, golden standard, Symmetry-Adapted Perturbation Theory (SAPT) interaction energies with errors well below chemical accuracy,\cite{giovannini2024kohn} with a substantial reduction of the computational cost. All KS-FEDA calculations were performed using a development version of the electronic structure code eT.\cite{eT_jcp}

\begin{acknowledgement}
This work was funded by the European Union – Next Generation EU in the framework of the PRIN 2022 PNRR project POSEIDON – Code P2022J9C3R. Computing facilities provided by CINECA HPC center (Iscra C project ``BHES'') are acknowledged.
\end{acknowledgement}

\begin{suppinfo}
Monophasic PMFs and density profiles, clustering statistics, and biphasic RDFs. Raw KS-FEDA data for all representative structures.  
\end{suppinfo}

{
\small
\bibliography{biblio}
}

\end{document}